\documentclass[twocolumn,prl]{revtex4}
\usepackage{graphicx,epsfig,amssymb}

\begin{document}

\title{Random matrix theory for closed quantum dots 
with weak spin-orbit coupling}
 
\author{K. Held$^*$, E. Eisenberg, and B. L. Altshuler}
\affiliation{Physics Department, Princeton University, Princeton, NJ 08544\\
 NEC Research Institute, 4 Independence Way, Princeton, NJ 08540}
\altaffiliation{ Present address: Max-Planck-Institut f\"ur Fest\-k\"or\-per\-for\-schung,
 70569 Stuttgart, Germany; k.held@fkf.mpg.de.}
\begin{abstract}
To lowest order in the coupling strength, the spin-orbit coupling 
in quantum dots results in 
a spin-dependent Aharonov-Bohm flux. This flux decouples 
the spin-up and -down random matrix theory ensembles of the quantum dot.
We employ this ensemble and find significant changes in the
distribution of the Coulomb blockade peak height,
in particular a decrease of the width of the distribution.
The puzzling disagreement between 
standard random matrix theory and the experimental distributions by 
Patel {\it et al.}\  
might possibly be attributed to these spin-orbit effects.
\end{abstract}


\date{\today}

\maketitle


The spin-orbit coupling in a two-dimensional semiconductor quantum well
mainly contributes through the  
Rashba \cite{Rashba} and Dresselhaus \cite{Dress} terms,
arising from the asymmetry of the confining potential 
and the lattice structure, respectively.
It is much weaker than in three-dimensional semiconductors
where it is induced mainly by impurities, which are absent in a high-mobility
two-dimensional electron gas. The spin-orbit scattering 
is further suppressed if the two-dimensional system
is confined to a quantum dot;
estimates of the spin-flip rates were given in Ref.\ \onlinecite{Khaetskii}.
This fact is of great importance for future applications
of quantum dots as spintronics devices. 
However, it was shown that spin-orbit scattering has
a significant effect in the presence of   
an in-plane magnetic field \cite{Halperin,Oreg,Khaetskii},
which explains \cite{Halperin} recent experiments
\cite{Folk}.

In this paper, we discuss another manifestation of the spin-orbit
coupling in confined structures, which takes place even  in the 
absence of appreciable spin-flip scattering.
Aleiner and Fal'ko recently showed \cite{Aleiner01a} 
that a  weak spin-orbit coupling
creates a spin-dependent Aharonov-Bohm flux. 
While this flux does not flip spins,
it can change the random matrix ensemble of the quantum dot. 
For broken time reversal symmetry,
the spin-up and spin-down parts of the spectrum are completely 
uncorrelated and described by independent Gaussian unitary ensembles (GUE)
\cite{Aleiner01a}. 
The possibility of such an ensemble was raised by
Alhassid \cite{Alhassid01}, while the relation to 
the spin-orbit coupling was already suggested by
Lyanda-Geller and Mirlin \cite{Lyanda}.
In the present paper, we study the statistical distribution
of the Coulomb blockade peak height in  this ensemble, 
and find the distribution to be narrowed.
This might explain the discrepancy between
a recent experiment by Patel {\it et al.} \cite{Patel98} and 
standard random matrix theory (RMT) \cite{Alhassid98,Alhassid01}
at low temperatures.

In Ref.~\onlinecite{Aleiner01a}, the free-electron Hamiltonian with
Rashba 
and Dresselhaus 
spin-orbit terms was expanded to second order
in the coordinates, under the assumption that
$L_{1,2}/\lambda_{1,2}\ll 1$ 
($L_{1,2}$: lateral dimensions of the two-dimensional quantum dot;
$\lambda_{1,2}$: characteristic length scale of the
spin-orbit coupling which is proportional to
the inverse spin-orbit coupling strength). One obtains
\begin{eqnarray}
\tilde{H} &=&\frac{1}{2m}\left( \vec{p}-e\vec{A}-\vec{a}_{\bot }\frac{\sigma
_{z}}{2}-\vec{a}_{\Vert }\right) ^{2}+u(\vec{r}). \label{Ham}
\end{eqnarray}
Here, $u(\vec{r})$ is the (disordered) confining potential;
$\vec{p}=\vec{P}-e\vec{A}$ is the kinetic momentum with the canonical momentum
$\vec{P}$ and the vector potential
\begin{eqnarray}
\vec{A} &=&B_{z}[\vec{r}{\bf \times }\vec{n}_{z}]/2c;\hspace{0.12in}\vec{a}%
_{\bot }=[\vec{r}{\bf \times }\vec{n}_{z}]/(2\lambda _{1}
\lambda_{2});\hspace{%
0.12in}  \label{aperp} \\
\vec{a}_{\Vert } &=&\frac{1}{6}\frac{[\vec{r}{\bf \times }\vec{n}_{z}]}{%
\lambda _{1}\lambda _{2}}\left( \frac{x_{1}\sigma _{1}}{\lambda _{1}}+\frac{%
x_{2}\sigma _{2}}{\lambda _{2}}\right)  \label{apar};
\end{eqnarray}
$\sigma_i$ denote the Pauli matrices and $B$ is the magnetic field 
in the direction [001] perpendicular to the lateral quantum dot.
The coordinates $x_1$ and $x_2$ are along the directions $[110]$ and 
$[1\bar{1}0]$ 
and we neglected the Zeeman splitting term as we are interested 
in the behavior at relatively low magnetic fields.
The term $\vec{a}_{\Vert }$ is responsible for spin-flips but it is of higher
order in the spin-orbit coupling strength than $\vec{a}_{\bot}$.
Thus, it  will be neglected in the following as we assume the spin-orbit 
coupling to be weak such that $\vec{a}_{\bot}$ dominates. 
The  $\vec{a}_{\bot}$ term has
exactly the same form as the vector potential $\vec{A}$
except for its spin-dependence. 
As an electron collects an  Aharonov-Bohm flux
on a close path due to the vector potential $\vec{A}$,
it also collects a spin-dependent flux due to  $\vec{a}_{\bot}$.
This spin dependent flux translates to a spin-dependent effective
magnetic field, so that the electrons feel a total magnetic field of strength
$B^{\rm eff}_{\sigma}= B+ \frac{c}{e}\frac{1}{\lambda_1 \lambda_2}
\frac {\sigma}{2}
$
with $\sigma=\pm\hbar$ for up- and down-spin, respectively.
An increase of the flux changes the matrix elements, and 
scrambles the eigenenergies and eigenvectors.
In the absence of spin-orbit coupling, the flux is exactly the same
 for spin-up and spin-down electrons such that their eigenenergies
and eigenvectors are degenerate. If the spin-orbit terms are present,
but no external magnetic field is applied, the time-reversal symmetry
is preserved, and the states are still Kramers 
degenerate (up-spin and
down-spin see the same magnitude of magnetic field with opposite signs).
However, when spin-orbit coupling and external magnetic field are present,
electrons with different spin see different magnetic fields, and 
their eigenenergies and eigenvectors decorrelate. 
If the spin-dependent flux is large enough
spin-up and spin-down eigenenergies and eigenvectors are distributed 
according to two {\em independent}  GUEs
\cite{Aleiner01a}.

\begin{figure}[tb]
\includegraphics[width=2.4in]{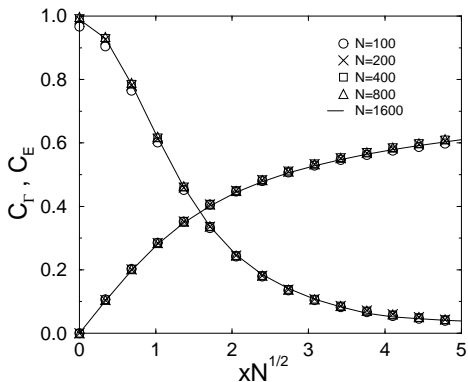}
\vspace{-.5cm}
\caption{Correlations of the level tunneling rates $C_\Gamma$ and
the rescaled spectral diffusion correlator $C_E$
for the RMT model (\ref{RMTmodel}), as a function of $x\sqrt{N}$
for different matrix sizes $N$.
The data collapse for different $N$ indicates the universality.
\vspace{-1.1em}} 
\label{figcorr}
\end{figure}

Before we analyze this weak spin-orbit RMT ensemble, we study the
decorrelation of the eigenenergies and
eigenvectors due to a change in the magnetic field, to determine how
much flux is needed in order to have two uncorrelated ensembles. 
The additional flux translates to a change of the random matrix
described by \cite{Aleiner01}
\begin{equation}
H= \frac{H_1+x H_2}{\sqrt{1+x^2}}
\label{RMTmodel}
\end{equation}
with RMT matrices $H_1$ and $H_2$ in the unitary ensemble.
As the perturbation $x$ increases from zero the eigenenergies $E_i(x)$
and eigenfunctions $\Psi_i(x)$ of $H$ 
change. We analyze  the decorrelations of the energies
via the  level diffusion correlator 
$C_E \!=\! \langle\!\langle\sqrt{(E_i(x)-E_i(0))^2  }\rangle\!\rangle /
\Delta$, 
where $ \langle\!\langle\cdots\rangle\!\rangle $ means
averaging over different realizations and different levels $i$. This
correlator has been shown to have a universal form \cite{Attias95}. The 
decorrelation of the eigenfunctions is measured by
$C_{\Gamma}\! =\! \langle\!\langle |\, \langle \psi_i(x)|\psi_i(0)\rangle
\,|^2 \rangle\!\rangle$. It can be shown that this correlator also measures
the correlations of the level tunneling rates
and has a universal form as well \cite{Alhassid96}.  
The results are presented in Fig.\ \ref{figcorr} and show that the 
correlations in both quantities disappear  
at about the same value of $x\sqrt{N}\! \approx\!  1$, where $N$ is the size of the random matrix.
Hence we conclude that the decorrelation of the eigenvalues
and the eigenfunctions (dot-lead coupling) occur together. 
Thus, the spin-orbit coupling leads to a
crossover from two degenerate GUE spectra, to an
ensemble of two uncorrelated GUE spectra.

For the above RMT model, the crossover to two independent GUE ensembles
occurs at $x\sqrt{N}\! \approx\!  1$. The corresponding flux difference needed
to decorrelate the spectrum is given by the following relation 
\cite{Aleiner01}
\begin{equation}
{x}\sqrt{N}=\chi\sqrt{g_T}\frac{\delta\Phi_{\rm eff}}{\Phi_0}
\label{estPhi}
\end{equation}
where $\delta\Phi_{\rm eff}$ is the flux difference, $\Phi_0$ is the quantum 
unit of flux, $g_T$ denotes the Thouless conductance, and $\chi$ is a 
non-universal sample-dependent constant of order unity.
We thus realize that one needs about $1/\sqrt{g_T}$ flux quanta 
to crossover to two uncorrelated GUE ensembles.

Let us now estimate the strength of spin-orbit interaction required
to create this amount of flux difference. As mentioned above, the difference
in effective flux between the two spin sectors is $\delta\Phi_{\rm eff}/\Phi_0=
L_1L_2/(\lambda_1\lambda_2)$.
The $\lambda$'s are connected to the Rashba and Dresselhaus spin orbit parameters $\gamma$ and $\eta$ via
$1/\lambda_1\lambda_2 =4(\gamma^2-\eta^2)$.
Independent estimates of $\gamma$ and $\eta$ are not available, but,
in principle, can be obtained \cite{Aleiner01a}. One can get an 
approximate value via the better known parameter
$Q_{\rm SO}^2=({\hbar v_F}/{E_F} )^2 (\gamma^2+\eta^2)$:
\begin{eqnarray}
\left | \frac{1}{\lambda_1\lambda_2} \right | & \leq &
Q_{\rm SO}^2 \left (\frac{2E_F}{\hbar v_F}\right )^2 = Q_{\rm SO}^2 k_F^2 \\
\frac{\delta\Phi_{\rm eff}}{\Phi_0}& = & Q_{\rm SO}^2 \, k_F L_1 \, k_F L_2
\end{eqnarray}
Typical experimental values are $k_F L_{1,2}\sim 50$,  $g_T\sim 10-100$,
and estimates for $Q_{\rm SO}$
are in the range $4-16 \times 10^{-3}$ \cite{Halperin}. Thus,
we estimate
 $\sqrt{g_T}\; \delta\Phi_{\rm eff}/\Phi_0=0.1-6.4$, i.e.,
the right hand side of Eq.~(\ref{estPhi}) can be expected to be
of order unity \cite{Zumbuehl02}. Hence,  the spin-orbit effect is
strong enough to decorrelate (or to start to decorrelate)
the spin-up and spin-down sector, while being
weak enough  not to yield significant spin-scattering.
A strong spin-scattering, which can be generated by
the application of an in-plane magnetic field,
would mix the two spin species and
result in a single GUE.

We will now analyze this situation where
the weak spin-orbit coupling results in  two uncorrelated
GUE ensembles for spin-up and spin-down electrons
and the results do not depend on
the spin-orbit strength.
%
%
Under the assumption of a constant Coulomb interaction
\cite{Aleiner01} and 
applying the Master equation for sequential tunneling through the 
quantum dot,  
the conductance of a closed quantum dot 
is given by    \cite{Beenakker91} (for a review see 
Ref.~\onlinecite{Alhassid01})
\begin{equation} 
\!G\!=\!\frac{e^2}{{\rm k_B} T} 
\!\! \sum_{i\sigma}
\!
\frac{\Gamma^L_{i \sigma} 
\Gamma^R_{i \sigma}}{\Gamma^L_{i \sigma}\!+\!\Gamma^R_{i \sigma}} 
P_{\rm eq}(N)P(E_{i \sigma}|N)[1\!-\!f(E_{i \sigma}\!-\!\mu)]. \label{Eq:G} 
\end{equation}
Here $\Gamma^{L(R)}_{i \sigma}$ is the tunneling rate between the $i$th  
one-particle eigenlevel of the dot with spin $\sigma$ and the left (right) 
lead, $E_{i \sigma}$ is the one-particle eigenenergy of this level,
$P_{\rm eq}(N)$ denotes the  equilibrium probability to find $N$ 
electrons in the dot (we assume the typical experimental situation
where the Coulomb blockade only allows $N$ and $N+1$ electrons in the
quantum dot),  
$P(E_{i \sigma}|N)$ is the canonical probability to have the $i$th level 
of the spin-$\sigma$ sector occupied given 
the  presence of $N$ electrons in the dot, and $f(E-\mu)$ is the Fermi function at 
the effective chemical potential $\mu$ which includes the charging energy. 
In Eq.~(\ref{Eq:G}), $\Gamma^{L(R)}_{i \sigma}$ is distributed
according to the Porter-Thomas distribution 
for the GUE
$P_{2}(\Gamma) = \frac{1}{\overline{\Gamma}} \exp(-\Gamma/
\overline{\Gamma})$,
which only depends on the mean value $\overline{\Gamma}$ of the distribution
(we assume this mean value to be the same for the coupling to the left
and right lead in the following).

At zero temperature, only one level ($i_1$,$\sigma_1$) contributes 
in Eq.~(\ref{Eq:G}) such that $\mu=E_{i_1 \sigma_1}$,
$P(E_{i \sigma}|N)=1$, and $P_{\rm eq}(N)=1/2$.
Thus, the zero temperature average conductance is given by
$\langle G\rangle = \frac{1}{12} \frac{\hbar \bar{\Gamma}}{k_B T} \frac{e^2}{\hbar}$ and the ratio of standard deviation to mean-value
becomes $\sigma(G)/\langle G\rangle=2/\sqrt{5}$.
Here, we have used $\langle{\Gamma^R_{i \sigma}}/{(\Gamma^L_{i \sigma}\!+\!\Gamma^R_{i \sigma})}  \rangle={1}/{3}$
and  $\langle{\Gamma^R_{i \sigma}}^2/({\Gamma^L_{i \sigma}\!+\!\Gamma^R_{i \sigma}})^2  \rangle={1}/{5}$ for the GUE distribution.
At low temperatures, there are a few  realizations of the
RMT eigenlevel distribution
where a second level  ($i_2$,$\sigma_2$) is within
an interval of order $k_BT$ around the first level at the Fermi energy.
Then, the second level also contributes to the 
conductance through the quantum dot. Neglecting the shift of the
chemical potential due to the second level 
(i.e., keeping  $\mu=E_{i_1 \sigma_1}$), we  calculated this two level
situation. This gives the leading behavior in $k_B T/\Delta$ for 
Eq.~(\ref{Eq:G}):
\begin{eqnarray}
\!\!\!\!\!\!\!\frac{\sigma(G)}{\langle G\rangle}&\!\!=\!\!&\frac{2}{\sqrt{5}}
\left(\!1\!+\!\left[\frac{781}{9} \ln 2\! -\!\frac{127}{3} \ln3\!-\!\frac{409}{27}\right]\frac{k_B
  T}{\Delta}\!\right)\!.
\label{Eq:lowT}
\end{eqnarray}

\begin{figure}[tb]
\vspace{-.5cm}

\vspace{-.5cm}
\includegraphics[width=2.8in]{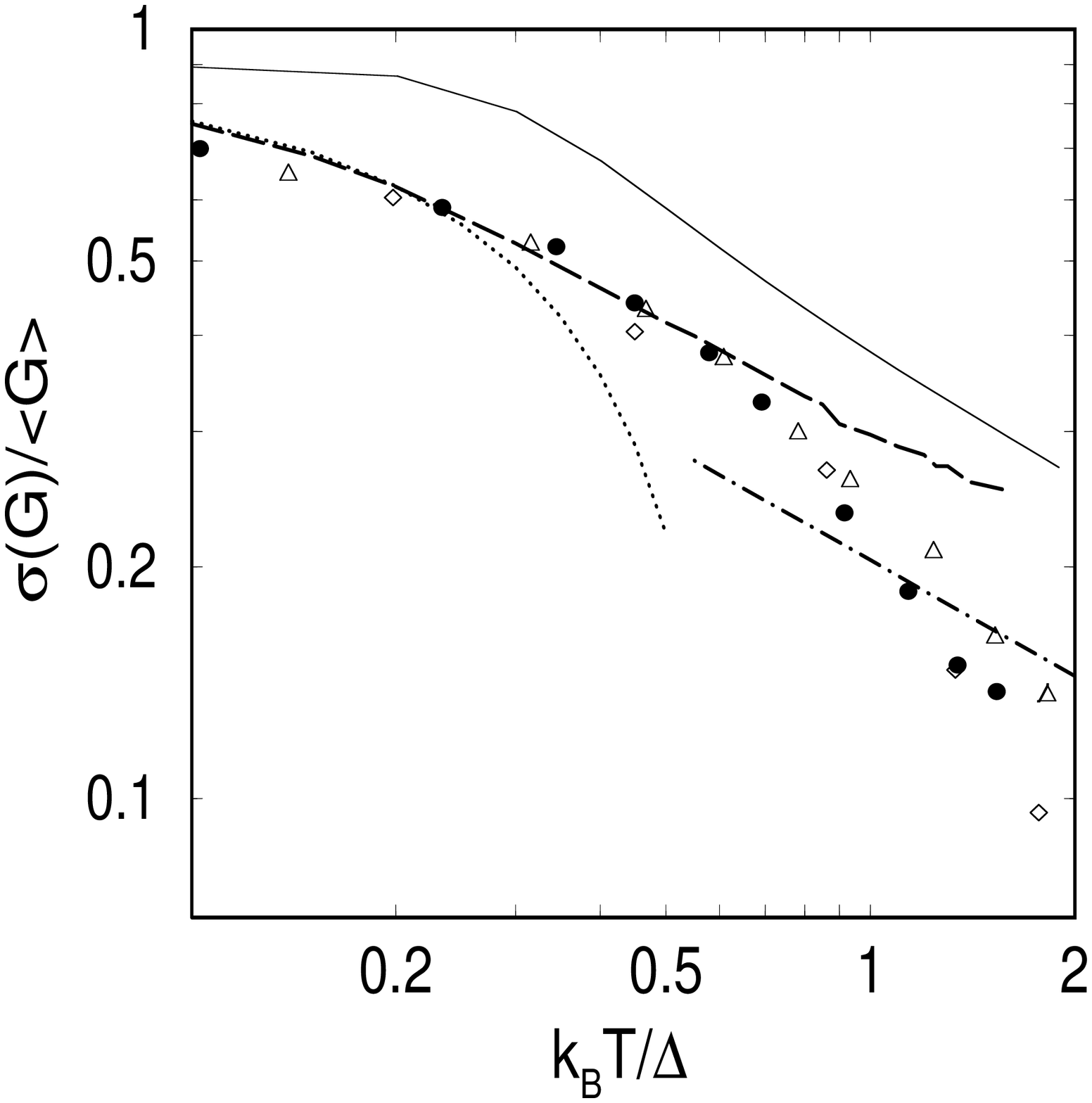}
\vspace{-2.99cm}
\caption{Width of the conductance distribution
$\sigma(G)/\langle G\rangle$ vs.\ temperature.
At low temperatures, 
the RMT ensemble for weak spin-orbit interaction [dashed line; 
dotted line: low temperature behavior
Eq.~(\ref{Eq:lowT})] well describes the experiment \cite{Patel98} (symbols correspond to slightly different quantum dots), in contrast to
standard RMT (solid line) \cite{Patel98}. At higher temperatures,
a further suppression is  due to
 inelastic scattering processes (dot-dashed line: $\Gamma_{\rm in}\!= \!\infty$ high-$T$ asymptote).
\vspace{-1.1em} }
\label{fig:Sigma}
\end{figure}

For general temperatures, 
we maximize numerically the conductance Eq.~(\ref{Eq:G}) w.r.t.\ $\mu$,
and averaged over 100000 RMT realizations
of the eigenenergies and the dot-lead couplings.
The results are shown in Fig.\ \ref{fig:Sigma}, in comparison to the
experiment of Patel {\it et al.}. 
In contrast to the standard RMT result \cite{Patel98},
the  RMT ensemble for weak spin-orbit coupling describes the width of
the conductance distribution and its change with
temperature reasonably well at low temperatures, 
without any adjustable parameter.
Compared to the standard GUE, the width of the distribution is
reduced at low temperatures because of
the absence of level repulsion for levels with
opposite spin. This results in higher probability to find a close-by level 
(with opposite spin and independent tunneling rate), and
leads to more RMT realizations in which two or more levels contribute
to the low-temperature conductance.
Having more independent channels for the conductance
makes the probability distribution more Gaussian and decreases its width.
At higher temperatures, the experimental results are not 
adequately described by
spin-orbit effects alone. In the   regime $k_B T \! \gtrsim \!\Delta$ however \cite{Eisenberg01}, one has to account for
inelastic scattering $\Gamma_{\rm in}$. 
Taking the limit $\Gamma_{\rm in}\! \rightarrow\! \infty$,
we obtain the high temperature asymptotic behavior
$\frac{\sigma(G)}{\langle G\rangle} \!=\!\sqrt{\frac{1}{24}\frac{\Delta}{k_BT}}$
which gives reasonable results except for the  quantum dot with diamond-symbols.
Note that upon reducing $k_B T/\Delta$,  $\Gamma_{\rm in}$ will
decrease, resulting in a crossover from the dot-dashed to the dashed line
in Fig.\ \ref{fig:Sigma}. The inelastic scattering rates of \cite{Eisenberg01} would imply that the dashed  $\Gamma_{\rm in}\!=\!\infty$-line
is approached in the range $k_B T$=1.5-4 $\Delta$.

\begin{figure}[tb]
\vspace{-.9cm}

\hspace{-.2cm} \includegraphics[width=1.8in]{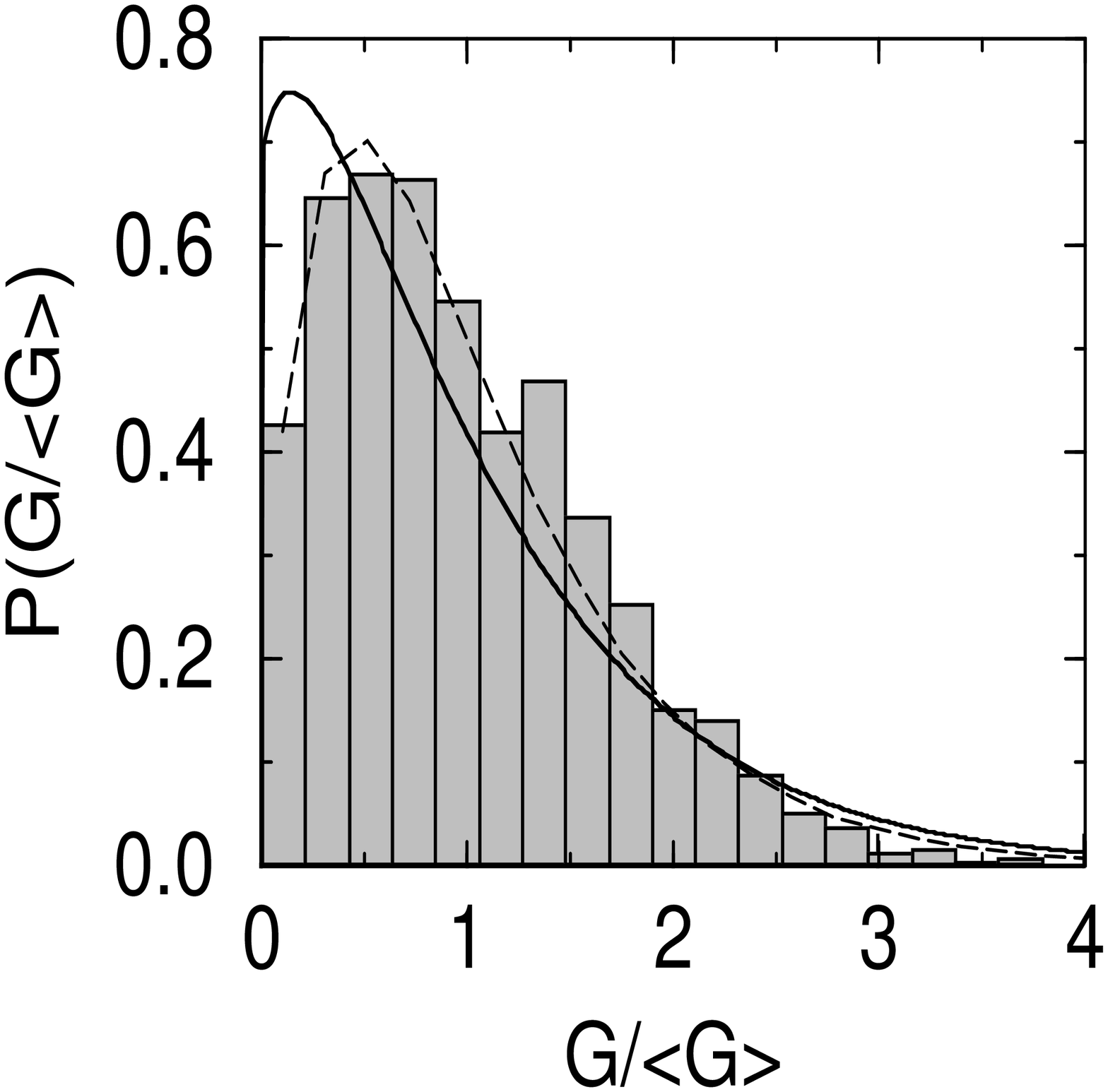}
\hspace{-.7cm} \includegraphics[width=1.83in]{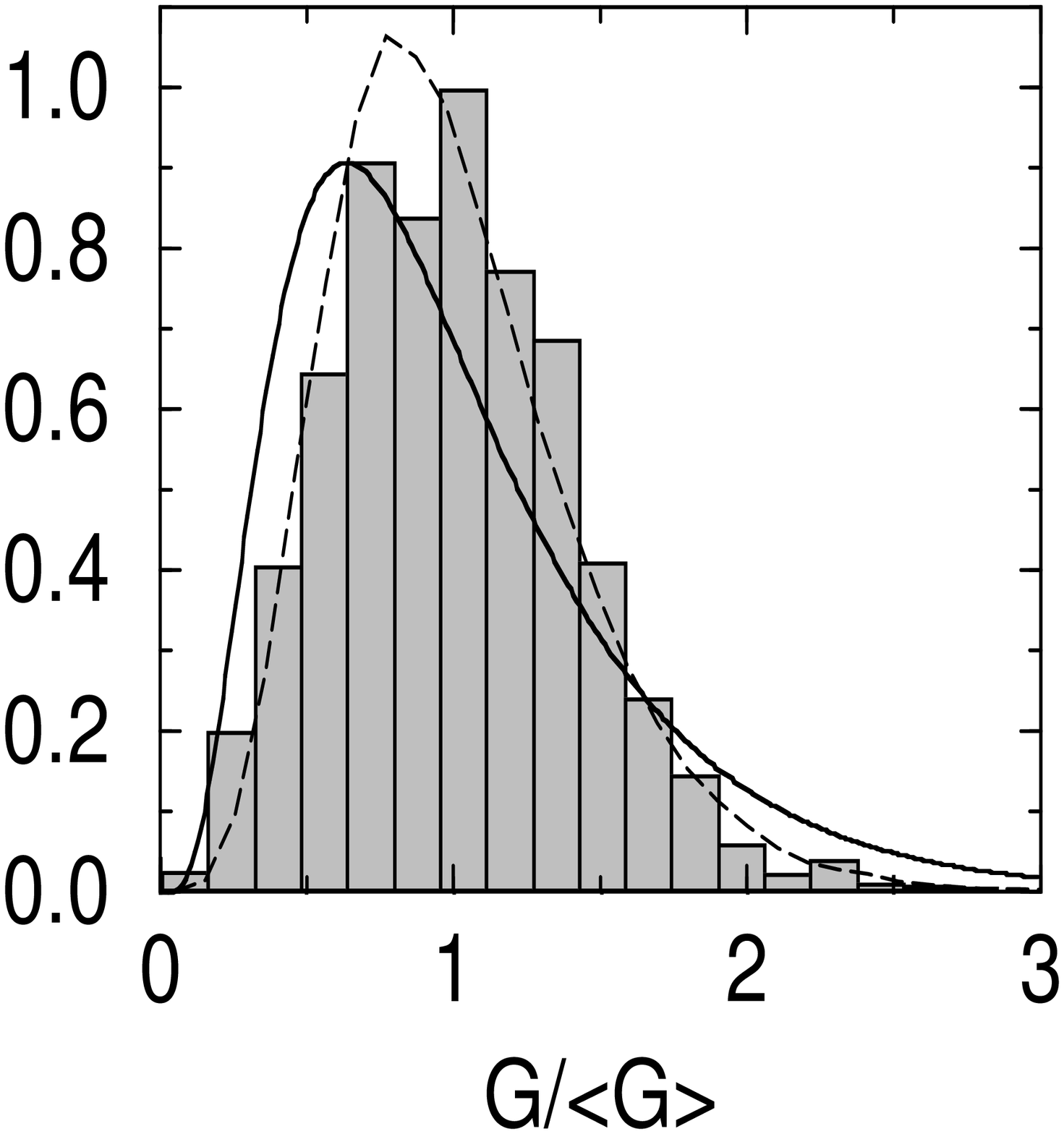}

\vspace{-1.cm}

\caption{RMT predictions with weak spin-orbit coupling (dashed line) for the probability distribution 
of the Coulomb blockade peak
conductance for a quantum dot at $k_B T=0.1\Delta$ (left figure) and
 $k_B T=0.5\Delta$ (right figure),
compared with the Patel {\it et al.} experiment \cite{Patel98} (histograms)
and standard RMT theory \cite{Patel98} (solid line).
There are no free parameters in these distributions.\vspace{-1.1em}}
\label{fig:Stat}
\end{figure}

In Fig.\ \ref{fig:Stat}, we compare the full probability
distribution with the experimental one~\cite{Patel98} at $k_B T=0.1\Delta$
and  $k_B T=0.5\Delta$.
Within the experimental statistical fluctuation, a  good
agreement is achieved without any free parameters, 
much better than for the standard RMT \cite{Patel98}.
This suggests that the spin-orbit strength is sufficient to fully 
decorrelated the spin-up and spin-down ensembles.
With an estimate of the experimental Thouless conductance $g_T \approx 20$ obtained from
$g_T\approx \sqrt{N}$, this means that a spin-orbit coupling strength
$Q_{\rm SO} \gtrsim  10^{-2}$ is required
in the quantum dot of Ref. \onlinecite{Patel98} 
(where we set $\chi=1$ in Eq.~(\ref{estPhi})).
In general, the crossover to the weak spin-orbit regime
occurs at $Q_{\rm SO}^2 (k_F L)^{5/2}\gtrsim 1$.
Thus, the size of the dot and the
parameter $Q_{\rm SO}$ which depends on the dot's specific
asymmetry of the confining potential  determine whether this quantum
dot is in the weak spin-orbit regime.
The size dependence might explain why earlier measurements 
by Chang {\it et al.} \cite{Chang96}
using very small quantum dots showed agreement with the standard
RMT without spin-orbit interaction. 
A similar agreement was found by Folk {\it et al.} \cite{Folk96},
despite using similar large quantum dots as in Ref.~\onlinecite{Patel98}.
The contradictory results of \cite{Folk96} and \cite{Patel98} might be 
due to the better statistics of the latter experiment,
or could be explained within the framework presented here, as following from
differences in the confining potential (which
might be, e.g., caused by differences in the realization of the 
two dimensional  electron gas and the gate voltage), 
translating into differences in $Q_{\rm SO}$. Alternatively, it is possible
that the spin-orbit effect in both samples is weak, and the 
deviations from RMT in \cite{Patel98} should be explained by another
mechanism (e.g. exchange \cite{UsajPC}).

In order to validate that the quantum dot is indeed in the weak 
spin-coupling regime described here, we suggest to repeat the experiment
with an in-plane magnetic field. A strong in-plane magnetic field
should drive the system towards the strong spin-orbit scattering limit.
In general, one would expect the spin-orbit scattering to suppress 
$\sigma(G)/\langle G\rangle$. However, in the case of weak spin-orbit coupling
the in-plane magnetic field, which drives the system towards a single
GUE,  regenerates the level repulsion.
Therefore, we predict  $\sigma(G)/\langle G\rangle$ to increase upon applying
an in-plane  magnetic field at low temperatures.
Another crucial test to the
the weak-spin orbit scenario is the behavior in the absence of
a
magnetic field.
In this case, the degeneracy is preserved but the spin-orbit coupling 
drives the system from the Gaussian orthogonal
to the unitary ensemble \cite{Aleiner01a}.
One implication is a strong suppression of  the magnetoconductance.

Finally we note that the disagreements between RMT predictions and the
results of \cite{Patel98} can not be attributed to
dephasing. Had this been the case, this experiment would indicate 
an appreciable dephasing even at low temperatures, in contradiction
with theoretical predictions \cite{Altshuler97}. However, 
recent measurements of the low-temperatures dephasing rates 
\cite{Folk00} are consistent with theory \cite{Eisenberg01}, and
furthermore, it has been shown by Rupp and Alhassid \cite{Rupp02} 
that dephasing alone can not explain the
results of \cite{Patel98}. Our calculation shows that
the spin-orbit coupling without dephasing can describe the 
low-temperature part 
of \cite{Patel98}, and that the inclusion of strong dephasing
gives reasonable agreement for the high-temperature part.

In conclusion, we analyzed the effect of weak spin-orbit coupling
on closed quantum dots in the presence of a perpendicular magnetic field
which breaks the time-reversal symmetry. In this regime which 
can be realized for (some) quantum dots, the  spin-orbit coupling
does not lead to one non-degenerate GUE ensemble but to
two independent GUEs for spin-up and -down
electrons. This has important consequences, 
in particular, at low temperatures, as there is no level-repulsion
for levels with opposite spins. The
statistical distribution of the conductance peak maximum
shows a good agreement with recent experimental distributions
by Patel {\it et al.}\ \cite{Patel98}, but disagrees
with experiments for similar sized quantum dots \cite{Folk96}. 
The exchange interaction might yield similar changes in the 
statistical distribution \cite{UsajPC}, and
it is unclear at present whether the complete explanation for the
peak heights statistics behavior is given by the weak spin-orbit RMT.
More experiments are needed
to clarify the relative importance of the two effects and to
explain the experimental contradiction mentioned above. 
If the spin-orbit effect is dominant, we predict
an increase of the width of the distribution upon applying a
strong in-plane magnetic field and a very low
magnetoconductance. We further note that  
without spin-degeneracy there will also
be {\em no} $\delta$-function-like 
contribution in the  level-spacing distribution,
in contrast to  standard RMT.

\begin{acknowledgments}
We acknowledge helpful discussions 
with I.L. Aleiner, Y. Alhassid, H.U. Baranger, C.M. Marcus, T. Rupp, and G. Usaj. 
This work has
been supported by ARO, DARPA, and the  Alexander von Humboldt 
foundation.
\end{acknowledgments}


\end{document}